%% file: main.tex
\documentclass[10pt, conference]{IEEEtran}
\usepackage{graphicx}
\usepackage{array, colortbl}
\usepackage{multirow}
\usepackage{amsmath}
\usepackage{svg}

\usepackage{rotating}
\usepackage{subfig}
\usepackage{float}
\usepackage{enumitem}
\usepackage[obeyspaces,hyphens,spaces]{url}
\usepackage{balance}
\usepackage[normalem]{ulem}
\usepackage{cleveref}

\usepackage{booktabs}

\usepackage{tcolorbox}

\newcommand{\rqone}{\textit{RQ1: What are the characteristics of time-to-first-response in the selected projects?}}
\newcommand{\rqtwo}{\textit{RQ2: How do bots affect the relationship between time-
to-first-response and the overall PR lifetime?}}
\newcommand{\rqthree}{\textit{RQ3: What are the characteristics of PRs with a longer
time-to-first-human-response? }}
\newcommand{\rqfour}{\textit{RQ4: What are the ratios of continuing PR contributors in PRs with short vs. long time-to-first-human-response? }}
\newcommand{\NA}{---}
\makeatletter
\renewcommand{\paragraph}[1]{\noindent\textsf{#1}.}

\title{Understanding the Time to First Response In GitHub Pull Requests}

\author{
    \IEEEauthorblockN{Kazi Amit Hasan, Marcos Macedo, Yuan Tian, Bram Adams, Steven Ding}
    \IEEEauthorblockA{School of Computing, Queen's University \\Kingston, ON, Canada
    \\\{kaziamit.hasan, marcos.macedo, y.tian, bram.adams, steven.ding\}@queensu.ca}
}

\begin{document}
\maketitle
\thispagestyle{plain}
\pagestyle{plain}
\begin{abstract}

The pull-based development is widely adopted in modern open-source software (OSS) projects, where developers propose changes to the codebase by submitting a pull request (PR). However, due to many reasons, PRs in OSS projects frequently experience delays across their lifespan, including prolonged waiting times for the first response. Such delays may significantly impact the efficiency and productivity of the development process, as well as the retention of new contributors as long-term contributors. 

In this paper, we conduct an exploratory study on the time-to-first-response for PRs by analyzing 111,094 closed PRs from ten popular OSS projects on GitHub. We find that bots frequently generate the first response in a PR, and significant differences exist in the timing of bot-generated versus human-generated first responses. We then perform an empirical study to examine the characteristics of bot- and human-generated first responses, including their relationship with the PR's lifetime. Our results suggest that the presence of bots is an important factor contributing to the time-to-first-response in the pull-based development paradigm, and hence should be separately analyzed from human responses. We also report the characteristics of PRs that are more likely to experience long waiting for the first human-generated response. Our findings have practical implications for newcomers to understand the factors contributing to delays in their PRs.
\end{abstract}


\section{Introduction} \label{sec:introduction}
\input{sections/introduction.tex}

\section{Related Work} \label{sec:related}
\input{sections/related.tex}

\section{Study Design} \label{sec:method}
\input{sections/method.tex}

\section{Preliminary Study Results} \label{sec:preliresults}
\input{sections/preliresults.tex}

\section{Empirical Study Results} \label{sec:results}
\input{sections/results.tex}

\section{Discussion}  \label{sec:discuss}
\input{sections/discuss.tex}

\section{Threats to Validity}  \label{sec:threats}
\input{sections/threats.tex}

\section{Conclusion} \label{sec:conclusion}

Delays in receiving the first response from a reviewer can decelerate the overall velocity of software development. In this work, we conduct a fine-grained analysis of the first stage in a PR's life, i.e., time-to-first-response, focusing on the impact of bots and the relationship between first human response time and PR lifetime as the future activities of contributors. We analyzed 111,094 closed PRs from ten popular and mature OSS GitHub projects. Our study shows several critical issues researchers and developers need to consider while analyzing and interpreting the time-to-first-response in PRs. The most important one is the need to separate first responses by bots and humans. We also identified different characteristics of PRs with a long time-to-first-human-response. 

In the future, we would like to expand the scope of analyzed GitHub projects and further analyze the intent of the first response and how it would impact the first response time. Moreover, we would like to consider other events in the pull requests and build a tool to predict the time to the first human response.

\section*{Acknowledgement}
We acknowledge the support of the Natural Sciences and Engineering Research Council of Canada (NSERC), [funding reference number: RGPIN-2019-05071].

\bibliographystyle{IEEEtran}
\bibliography{main.bib}
\end{document}

%% file: sections/introduction.tex
Modern open-source software (OSS) development has been embedded in a collaborative and distributed environment, allowing development teams to receive contributions from external developers interested in collaborating on a particular project. Pull-based development has become a common paradigm for managing contributions to OSS projects~\cite{moreira2021factors}. Leveraging pull requests, contributors can independently modify project artifacts and subsequently request the merge of their proposed changes into the mainline development branch. A pull request (PR), including bug fixes or new features, may be accepted and merged following assessment by a core team member, or rejected. Before getting its final decision, a PR has to pass through multiple stages, including waiting for the reviewer stage, review stage, and the final potential integration stage, any of which can be delayed due to inactive engagement of PR authors, triagers, reviewers, and integrators, as well as many other factors. 

Recently, Zhang et al.~\cite{zhang2022pull} reported that many factors explain the delay in the whole lifetime of a PR. They found that the time from the creation of a PR until its first comment, i.e., \textit{time-to-first-response}\footnote{Because we considered multiple types of first response in this study, we use the term \textit{time-to-universal-first-response} in the rest of this paper to refer to the time-to-first-response concept used in prior studies.}, plays the most important role in influencing PR latency if there exist comments in a PR. However, the characteristics of time-to-first-response remain unknown. 

To fill this gap, in this study, we perform a fine-grained analysis on the time-to-first-response in GitHub PRs. There are several reasons for studying the latency of this stage. First, besides the important role of this value in PR lifetime prediction, researchers also reported that time-to-first-response highly correlates with the PR outcome~\cite{yu2015wait,zhang2022pull2}. Second, understanding time-to-first-response could provide insights for developing PR-accelerating tools like Microsoft's Nudge~\cite{maddila2020nudge} on OSS projects. Nudge is a tool 
that could notify developers to interact with a pull request if there is not enough activity and its lifetime is past the predicted one. 
Third, new pull-based development practices, especially the adoption of software development bots, may impact the analysis of the PR latency in different stages, but its impact is still unknown. Our study would shed light on future research on the fine-grained latency of GitHub PRs and support the development of tools that could optimize the pull-based development process. 

To better understand the basic statistics of time-to-first-response in GitHub PRs, we first conduct a preliminary study leveraging the most recent and largest PR latency benchmark dataset provided by Zhang et al.~\cite{zhang2022pull}, which consists of 3,347,937 closed PRs from 11,230 GitHub projects. We observe that the time-to-first-response should be carefully interpreted as many PRs with extremely short (less than one minute) time-to-first-response are generated by bots. Therefore, we propose to redefine this concept by considering whether the response is generated by a human or bot (see Table~\ref{tab:terms}). Then, we conduct an in-depth empirical study of various types of time-to-first-response in ten popular and mature projects to answer the following four research questions:

\begin{itemize}[leftmargin=*]
        \item[] \textbf{\rqone} In this RQ, we report descriptive statistics of three types of time-to-first-response, i.e., the \textit{time-to-universal-first-response}, the \textit{time-to-first-human-response} and the \textit{time-to-bot-first-response}. The universal-first-response is the first pull-request comment (PR comment) or code review comment not given by the PR author, i.e., either by a bot or a human. The time-to-first-human-response is the time from PR creation until the first human response, which would be longer than the universal-first-response if a bot-generated response existed before the first human response. Therefore, comparing the two values would indicate the impact of bots on the interpretation of time-to-first-response. The time-to-bot-first-response statistics would indicate how fast the bot-generated first response appears.
    

    \vspace{0.1cm}
    \item[] \textbf{\rqtwo} Zhang et al.~\cite{zhang2022pull} has reported that the majority (58.7\%) of the variance in PR lifetime can be explained by the time-to-universal-first-response for PR with comments. However, since the universal first response could be generated by two different types of accounts, i.e., bot or human, it is unknown if the pattern still holds when considering only human responses or bot responses. Answering this question would help developers determine which type of first response they should use as the indicator of PR lifetime.
    
    
    \vspace{0.1cm}
    \item[] \textbf{\rqthree} This RQ aims to analyze the factors behind long human response time within an OSS project. Here we did not consider the universal-first-response-time and bot-response-time as the bot's response time can be easily controlled. We answer this RQ using both quantitative and qualitative methods~\cite{gregar1994research}.
    
     \vspace{0.1cm}
    \item[] \textbf{\rqfour} We hypothesize that the time-to-first-human-response might affect the attitude of contributors toward future contributions. In this RQ, we report the ratio of newcomers and existing PR contributors continuing to contribute PRs in two types of PRs, i.e., PRs that received the first human response within one day, and more than one day.

\end{itemize}

To answer the above RQs, we leverage the PR latency benchmark data and extend it by collecting more PR information from the target ten projects. We develop a new approach to determine if a PR response is generated by a bot or human based on the BoDeGha tool~\cite{Golzadeh2021JSS}, and manually evaluate the accuracy of this approach by examining 1,000 PRs.

\vspace{0.1cm}
\noindent \textbf{Main contributions:} We conduct the first exploratory study on the time-to-first-response in GitHub PRs. Our results suggest future researchers and practitioners who want to leverage time-to-first-response to (1) be aware that the first response in a PR can be generated by a bot. (2) Time-to-first-human-response is still a good indicator for PR lifetime estimation in all PRs with human responses. (3) Less complex PRs and better communication between PR authors and core developer team members correlate with shorter time-to-first-human-response. (4) The time-to-first-human-response correlates with the future PR contribution likelihood of newcomers. 

\noindent \textbf{Replication package:} We provide a replication package with all of our data and scripts.\footnote{\url{https://doi.org/10.5281/zenodo.7747317}}


%% file: sections/related.tex
\begin{table*}[t]
\caption{Definition of terms used in this paper.}
\label{tab:terms}
\centering
\small
\begin{tabular}{>{\itshape}ll}
\toprule 
\textup{Term}       & Definition\\
\midrule
Response                 & A PR comment or a code review comment within a PR.\\
Human Response                     & A PR comment or a code review comment given by a human other than the PR author.\\
Bot Response                    & A PR comment or a code review comment given by a bot.\\
Bot-first PR                     & A PR is a bot-first PR if its first response is given by a bot.\\
Human-first PR                    & A PR is a human-first PR if its first response is given by a human other than the PR author.\\
(Time-to-)universal-first-response & The (time from PR creation to the) first response in a PR.\\
(Time-to-)first-human-response & The (time from PR creation to the) first human response in a PR.\\
(Time-to-)bot-first-response & The (time from PR creation to the) first response (given by bot) in a bot-first PR.\\
\bottomrule
\end{tabular}
\end{table*}

\subsection{Time to First Response in Code Review}
In the prior studies on code review~\cite{bird2015lessons,macleod2017code, kudrjavets2022mining}, time-to-first-response is defined as the time from the publication of a code change until the first acceptance, comment, inline comment (comment on specific code fragments), or rejection by a person other than the author of the code. Following this definition, bot-generated responses should not be considered. However, we observe that not all research has considered this when creating the time-to-first-response feature for their analysis on PR latency and decision~\cite{zhang2022pull, zhang2022pull2}. As such, it is still unknown how bots would affect our interpretation of time-to-first response. 


The closest related work to our research is by Kudrjavets et al.~\cite{kudrjavets2022mining}. They analyzed the waiting times between acceptance and merging in the code review process by mining data from two code collaboration platforms, i.e., Gerrit~\cite{Gerrit} and Phabricator~\cite{Phacility}. They found that non-productive waiting time could also exist while waiting for the first response. 
Our analysis goal is different from theirs. First, we target GitHub PRs, the structure of which is different from those on code review platforms. Secondly, we define three types of time-to-first-response and discuss their differences and relationship with PR lifetime. Last but not least, we explore the characteristics of PRs with longer time-to-first-human-response and future PR contribution activities of authors who received their first human response in different time ranges. The results of our study would complement their findings. 


\subsection{Other Studies on Pull Requests}

Besides the PR-related studies conducted by Zhang et al.~\cite{zhang2022pull,zhang2022pull2} (see Section~\ref{sec:introduction}), other studies on PRs mainly focus on understanding how the characteristics of GitHub PRs influence their acceptance~\cite{gousios2014exploratory,gousios2016work} and latency in review process~\cite{yu2015wait}, or both~\cite{pinto2018gets}. Researchers also investigated the PRs that have or would waste time and effort of developers' contributions and reviews, i.e., duplicated PRs~\cite{li2020redundancy} and abandoned PRs~\cite{khatoonabadi2021wasted} in GitHub projects. For instance, Khatoonabadi et al.~\cite{khatoonabadi2021wasted2} found that abandoned PRs tend to be more complex, their contributors tend to be less experienced, and their review processes tend to be lengthier than non-abandoned PRs. Similar to the above studies, our study explores the characteristics of PRs, but with a different focus, i.e., we split PRs based on their time-to-first-response characteristics. 

Researchers also build tools to support the management and analysis of PRs, including PR-accelerating tools like Microsoft Nudge~\cite{maddila2020nudge}, and duplicate PR detection tools~\cite{li2021detecting,wang2019duplicate}. Golzadeh et al.~\cite{Golzadeh2021JSS} propose a tool named BoDeGha which can identify bots from PRs by analyzing pull requests and issue comments. In this project, we leverage BoDeGha to collect the candidate bots and manually verify the results (see Section~\ref{sec:method}). 


%% file: sections/method.tex
\subsection{Benchmark Dataset }
This study is based on the recent, largest PR latency benchmark dataset created by Zhang et al.~\cite{zhang2022pull}. This dataset comprises 63 features extracted from 3,347,937 closed pull requests (meaning a decision has been made) in 11,230 OSS projects on GitHub, mined from the June 1st, 2019 GHTorrent MySQL data dump.\footnote{http://ghtorrent-downloads.ewi.tudelft.nl/mysql/mysql-201906-01.tar.gz} As our study focuses on the time-to-first-response, we exclude all PRs without this feature value, i.e., PRs without any comments. The filtered data contains 1,613,224 pull requests from 10,374 projects. 


\subsection{Studied Projects}

To answer RQ1-4, we selected ten projects based on the rich history of pull request data and the popularity of projects measured by stars. Specifically, we selected the top-10 projects with more than 1,000 closed PRs and more than 100 stars. A similar approach has been considered in the literature~\cite{yu2015wait, han2019characterization}. The background of each selected project is shown in Table~\ref{tab:list of projects}.


\begin{table*}[h]
\centering
\caption{Basic information of the ten selected projects. }
\label{tab:list of projects}

\begin{tabular}{{llllll}}
\toprule
 \textup{Project}        & Creation Dates of PRs &  \#PRs  & \#Stars & Domain  & Programming Language   \\ \hline
          Ansible                  & 2012-03 to 2019-05    & 15,387        & 39,616          & Automation Platform                                                                     & Python      \\ 
         Spark                       & 2014-03 to 2019-05      & 20,806       & 23,331     & Data Analytics Engine                                                                & Scala, Python   \\ 
          Kubernetes                & 2017-04 to 2019-05       & 18,663      & 45,686          & Container Orchestration                                                                 & Go       \\
 Rails                     & 2010-09 to 2019-05    & 12,707  &         48,587          & Web Application Platform                                                                & Ruby           \\ 
Salt                     & 2011-05 to 2019-05    & 10,127 &          10,869          & Automation Platform                                                                     & Python     \\ 
Odoo                     & 2014-05 to 2019-05   & 9,776  &          14,666          & Business Apps                                                                           & JavaScript, Python     \\ 
Node                 & 2014-11 to 2019-05  & 8,418   &         67,208          & JavaScript Runtime Environment               & JavaScript             \\ 
 Akka                            & 2011-05 to 2019-05  & 6,578	           & 10,717          & Application Development Toolkit              & Scala         \\ 
Discourse                & 2013-02 to 2019-05   & 4,319          & 30,508          & Community Discussion Platform               & Ruby       \\ 
 Hazelcast               & 2012-04 to 2019-05   & 4,313           & 3,358           & Distributed Computation and Storage Platform & Java        \\ \bottomrule
\end{tabular}
\end{table*}

\subsection{Additional Data Collection}
The benchmark dataset contains neither the authors of (universal-)first-responses, nor responses - there is no way to identify whether a bot or a human gives the first response. Therefore, we recollect all the PR comments, code review comments, and creation dates for the selected set of PRs, which are used to generate three new features, i.e., \textit{time-to-universal-first-response}, \textit{time-to-first-human-response}, \textit{time-to-bot-first-response}. Also, for RQ4, to determine if a PR author in our PR dataset continues to create PR(s), we collect for the 10 selected projects the authorship information for the developers of all PRs created before June 1st, 2020. This date falls one year after the end date of our benchmark dataset (June 1st, 2019) 
to guarantee we have at least one year to capture the future activities of authors who only contribute at the end of the considered data set. All the additional data to the benchmark dataset was collected automatically using a script, via the official GitHub v4 API.\footnote{https://docs.github.com/en/graphql}




\subsection{Pull Request Features}
In RQ2 and RQ3, we perform analysis based on PR features from three dimensions, i.e., pull request, project and developer. 


In RQ2, we follow the same methodology as Zhang et al.~\cite{zhang2022pull} to analyze the relationship between time-to-first-response and PR lifetime. We considered the same set of features available at PR closing time for PRs with comments, with some changes. Specifically, we added two new features: \textit{time-to-bot-first-response} and \textit{time-to-first-human-response} (ref. Table~\ref{tab:terms}) to understand their relationship with PR lifetime. Additionally, we excluded the \textit{ci\_exist} feature as, during our manual examination on PRs, we found that the values of the labels are inaccurate on our selected projects. The feature \textit{ci\_exist} indicates whether the PR contains a CI tool, as identified using a heuristic approach~\cite{zhang2022pull}. Lastly, to account for multicollinearity, we removed the feature \textit{sloc} that had a Variance Inflation Factor (VIF) greater than 5, i.e., this feature can be represented by other features.

In RQ3, we select all PR features available at PR opening time (see Table~\ref{features}), i.e., available before the universal-first-response and first-human-response. We leverage them to analyze the characteristics of PRs with longer time-to-first-human-response.




\begin{table*}[h]
 \centering
 \caption{Factors that explain a pull request's time-to-first-response on GitHub.}
 \label{features}
\begin{tabular}{{>{\itshape}cll}}
\toprule
\textup{Dimension}  & Feature                  & Description                                                                                                         \\ \hline
Pull Request       & description\_length               & length of pull request description                                                                                           \\ \cline{2-3} 
                   & first\_pr                         & first pull request? yes/no                                                                                                   \\ \cline{2-3} 
                   & hash\_tag                         & “\#” tag exists? yes/no                                                                                                      \\ \cline{2-3} 
                   & at\_tag                           & “@” tag exists? yes/no                                                                                                       \\ \cline{2-3} 
                   & num\_commits\_open                & \# of commits at pull request open time                                                                                      \\ \cline{2-3} 
                   & files\_changed\_open              & \# of files touched at pull request open time                                                                                \\ \cline{2-3} 
                   & src\_churn\_open                  & \# of lines changed (added + deleted) at pull request open time                                                              \\ \cline{2-3} 
                   & test\_churn\_open                 & \begin{tabular}[c]{@{}c@{}}\# of lines of test code changed test  (added + deleted) at pull request open time\end{tabular} \\ \cline{2-3} 
                   & churn\_addition\_open             & \# of added lines of code at churn request open time                                                                         \\ \cline{2-3} 
                   & churn\_deletion\_open             & \# of deleted lines of code at pull request open time                                                                        \\ \cline{2-3} 
                   & commits\_on\_files\_touched\_open & \# of commits on files touched at pull request open time                                                                     \\ \hline
Project            & team\_size                        & \# of active core team members  in the last three months   \\ \cline{2-3}                                                                 
                   & project\_age                      & \# of months from project to pull request creation                                                                           \\ \cline{2-3} 
                   & open\_pr\_num                     & \# of open pull requests                                                                                                     \\ \cline{2-3} 
                   & sloc                              & executable lines of code                                                                                                     \\  \cline{2-3} 
                   & perc\_external\_contribs          & \% of external pull request contributions                                                                                    \\ \cline{2-3} 
                   & integrator\_availability          & latest activity of the two most active integrators                                                                           
                   \\ \hline
Developer          & prior\_review\_num                & \# of previous reviews in a project                                                                                          \\ \cline{2-3} 
                   & same\_affiliation                 & same affiliation contributor/integrator? yes/no                                                                              \\ \cline{2-3} 
                   & core\_member                      & core member? yes/no                                                                                                          \\ \cline{2-3} 
                   & prev\_pullreqs                    & \# of previous pull requests                                                                                                 \\ \cline{2-3} 
                   & social\_ strength                  & fraction of team members interacted with in the last three months                                                            \\ \cline{2-3}
                   & same\_user                        & same contributor and integrator? yes/no                                                                                      \\ \cline{2-3} 
                   & followers                         & \# of followers at pull request creation time                                                                                \\ \bottomrule
\end{tabular}
\end{table*}

\subsection{Bot Identification and Data Verification}
 We used a tool named BoDeGHa~\cite{Golzadeh2021JSS} to identify candidate bots that may appear in the PRs of selected projects. We picked this tool as it is a recent tool reported to perform well and can be directly applied on a given GitHub repository. However, while manually checking the returned identified list of bots and humans by BoDeGHa, we found that many core developers are identified as bots. Therefore, we developed a semi-automatic approach to identify the list of bots that may appear in our collected PRs, as follows:



\begin{itemize}
    \item Step 1: Run BoDeGHa on the selected 10 projects to collect an initial set of candidate bots.
    \item Step 2: Manual Verification :
    \begin{itemize}
        \item Manually review the profiles and activity (including comments, events, and commits) in the relevant repository for all accounts flagged as bots by BoDeGha. Any human accounts identified as bots were eliminated.
        
        \item Manually search for potential missing bots in the list of humans identified by BoDeGha using the \textit{``bot, robot, auto''} keywords. Read their profiles and add missing bot names to the bot list.
    \end{itemize}
\end{itemize}



The above heuristics may still fail to capture bots and lead to wrong information on different types of time-to-first-response. Therefore, we manually validated that the first-response related information (bot-first vs. human-first response, response dates) is correct. We randomly sampled 100 PRs from each of the 10 projects, for a total of 1,000 PRs. Then, we set up Bernoulli trials \cite{papoulis2002probability} with a \textit{success} being that all the examined information is correct and a \textit{failure} that some of it is not. We examined each PR by browsing through its public GitHub URL and counted the number of \textit{successes}. There were 998 \textit{successes} over the 1,000 PRs examined - 0.2\% inconsistent cases are observed.

In the end, we filter out PRs without human response as we compare time-to-first-human-response in both bot-first PRs and human-first PRs. The resultant number of PRs for each selected project is shown in Table~\ref{tab:list of projects}.




\subsection{Statistical Analysis Method}
In RQ1 and RQ3, we apply the following two types of statistical tests to compare different types of time-to-first-response and PR characteristics. \textbf{We consider the overall result of the experiment to be significant if it is statistically significant and the effect size is not negligible.}
\begin{itemize}
    \item To test the statistical difference between two comparison targets, we apply the Mann–Whitney test~\cite{mann1947test}. To determine if the difference is significant, we calculated the adjusted significance levels to account for multiple testing using the Holm-Bonferroni correction, following the suggestion by literature work~\cite{kitchenham2002preliminary}. Specifically, we use the \textit{multipletests} function from Python library \textit{statsmodels.stats.multitest}.


    \item To further test the practical difference between two targets with significant differences, we use Cliff’s delta to estimate their magnitude of difference (i.e., effect size). The value of Cliff’s delta (denoted by d) ranges from $-1$ to $+1$. In order to make comparisons more manageable, we transform the d values to qualitative magnitudes based on the following thresholds, which were provided by Hess and Kromrey~\cite{khatoonabadi2021wasted,hess2004robust}.

\end{itemize}
\vspace{0.1cm}
\[
    \text{Effect Size} = 
    \begin{cases}
        \text{Negligible} & \text{if $|d| \leq 0.147$,}\\
        \text{Small} & \text{if $0.147 < |d| \leq 0.33$,}\\
        \text{Medium} & \text{if $ 0.33 < |d| \leq 0.474$,}\\
        \text{Large} & \text{if $ 0.474 < |d| \leq 1$,}\\
    \end{cases}
\]
\vspace{0.1cm}

%% file: sections/preliresults.tex



To understand the basic statistics of time-to-universal-first-response in the PR latency benchmark, we split the lifetime of a PR into two stages by the (universal-)first-response, i.e., the \textit{time-to-first-response stage} and the \textit{after-first-response stage}. For each stage, we define four intervals. This section uses the terms ``first response'' and ``universal first response'' interchangeably.

Figure~\ref{fig:FTR_benchmark} shows the number of PRs in each time interval in the time-to-first-response stage. We observe that:

\begin{figure}[t]
    \centering
    \includegraphics[width=\linewidth]{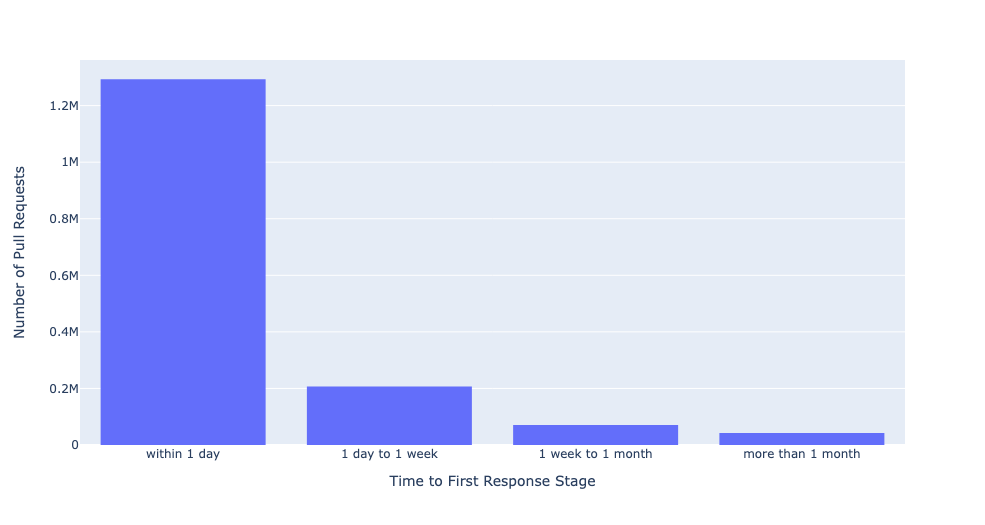}
    \caption{Number of pull requests in each category of time-to-universal-first-response.}
    \label{fig:FTR_benchmark}
\end{figure}

\textbf{80.16\% of the pull requests receive their universal first response within less than a day.} 68.36\% of pull requests even received their first response within an 8-hour workday. Furthermore, 93.01\% of the PRs received their first response within one week. 4.38\% of the PRs waited for one week to one month for their first response, and the remaining 2.61\% PRs waited more than one month for the first response.

 \textbf{39.78\% of the pull requests with same-day first response receive this response within 10 minutes}. 
 To understand why those PRs responded so fast, we conducted a manual study by randomly sampling 50 PRs with time-to-first-response less than or equal to 10 minutes. \textbf{We found that 35/50 (70\%) of the cases have a first response generated by a bot}. We observed CLA (Contributor License Agreement) bots, review supporting bots, and reviewable services.\footnote{https://reviewable.io/}
 

\begin{figure}[t]
    \centering
    \includegraphics[width = \linewidth]{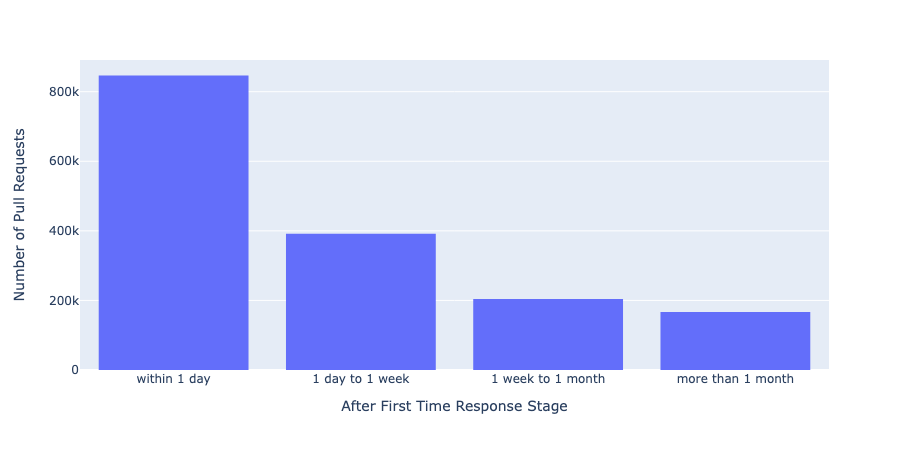}
    \caption{Number of pull requests in each category of time-after-first-response. \textit{Time-after-first-response} indicates the duration of the remaining PR lifespan after the universal first response.}
    \label{fig:AFTR_all}
\end{figure}

\begin{figure}[t]
    \centering
    \includegraphics[width=\linewidth]{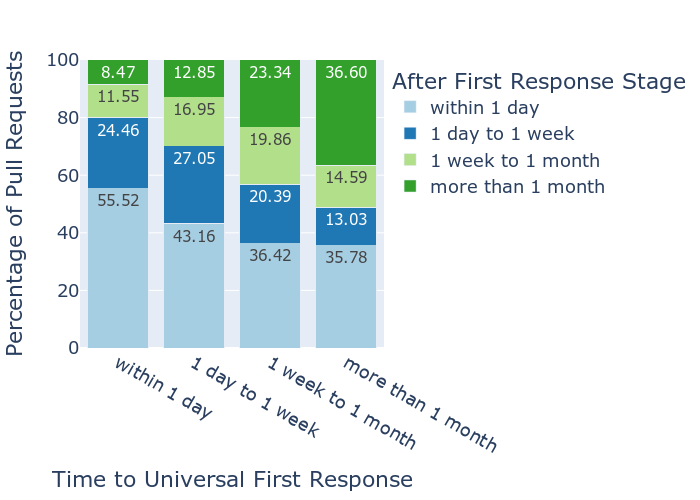}
    \caption{Percentage of each time-after-first-response category for each time-to-first-response category. }
    \label{fig:stackedbar}
\end{figure}

Furthermore, we also count the number of PRs across four time intervals for the after-first-response stage (ref. Fig.~\ref{fig:AFTR_all}). We found that 52.62\% and 76.96\% of PRs complete the rest of their lifetime within one day and one week respectively. 23.04\% PRs take more than one week to close after receiving the first response.

 

\textbf{Despite receiving their first response after weeks, a non-trivial proportion of PRs are quickly reviewed and merged.} We observe that 35.78\% of the PRs, which took more than one month to receive the first response, took another less-than-one day to complete their life (ref. Fig.~\ref{fig:stackedbar}). A similar percentage of PRs can be observed within PRs that receive their first response between one week and one month. To understand the potential reasons behind the long time-to-universal-first-response and shorter after-first-response stage time, we randomly sampled 50 PRs that waited more than one month for the first response and then closed within one day. We find that: 
    \begin{itemize}
        \item In 37 (74\%) PRs, the reviewer responded after one week but quickly merged the PR after reviewing it. In some cases, the authors had to grab the attention of the reviewer by mentioning their PRs in the comment of other PRs. \textbf{If a first response would arrive earlier, these cases might have a significantly shorter lifetime. }
        \item 10 (20\%) PRs were incomplete upon initial submission. Reviewers had to wait and thus start late.
        \item The remaining 3 cases are hard to interpret.
    \end{itemize}

    


\begin{tcolorbox}
\textit{Preliminary Study Summary:} Bots are responsible for most of the quickest (universal-)first-responses in our examined PRs. Out of the PRs that experience a long time to first response, 36\% of them have a short review process. Those pull requests' lifetimes might be shortened if they get their first response earlier.
\end{tcolorbox}
\vspace{0.1cm}


%% file: sections/results.tex
\subsection{\rqone} \label{sec:rq1} 

 
\noindent \textbf{Approach:} Following the terminology in Table~\ref{tab:terms}, we consider five types of first response time based on the response author type (bot or human) and PR type (bot-first, human-first or universal), i.e., universal first response, first human response, first human response in human-first PRs, first human response in bot-first PRs and first bot response in bot-first PRs. For an in-depth analysis of the characterization of time to first response, we formulated three sub-research questions to answer RQ1. 

\subsubsection{What are the statistics of the universal first response time versus first human response time?}

Previous studies~\cite{zhang2022pull} only reported the universal first response as they did not consider the author type (human or bot) of the response. In this sub-question, we compare the universal first response time with the first human response time in all PRs. Answering this question would reveal the differences between these two types of time to first response and guide practitioners in future analysis. 
The statistical comparison results between \textit{time-to-universal-first-response} and \textit{time-to-first-human-response} are shown in Table \ref{des_stats_RQ1,1}.

\vspace{0.1cm}

\noindent \textbf{Results:} \textbf{Universal first response is, in general, fast (within a day), in all selected projects.} The median time-to-universal-first-response ranges from 0.08 to 86.65 minutes, excluding Odoo. Odoo demonstrates a substantially longer median response time, i.e., 1413.17 minutes, compared to other projects, and its bot-first PR rate is only 0.69\%. Another extreme case we observed is in Discourse, where the median universal response time is a mere 0.08 minutes, with 97.08\% of its PRs being bot-first PRs.


\begin{table*}[]
\centering
\caption{Descriptive statistics about the ten selected projects. Time-to-universal-first-response (TUFS) and time-to-first-human-response (TFHR) are given in minutes. \#Bots = number of unique bots detected in bot-first PRs, M = mean, Mdn = median, Std = standard deviation.}
\label{des_stats_RQ1,1}
\scalebox{1.1}{ 
\begin{tabular}{{lcc|ccc|ccc}}
\toprule
\begin{tabular}[c]{@{}c@{}}Project \end{tabular} & \#Bots & \begin{tabular}[c]{@{}c@{}}\% of Bot-first PRs\end{tabular} & \textbf{}                       & TUFS & \textbf{}    & \textbf{}                       & TFHR & \textbf{}    \\ \cline{4-9} 
\textbf{}                                                        & \textbf{}     & \textbf{}                                                            & \multicolumn{1}{c|}{M} & \multicolumn{1}{c|}{Mdn} & Std & \multicolumn{1}{c|}{M} & \multicolumn{1}{c|}{Mdn}  & Std \\ \hline
Akka                                                             & 2             & 56.63\%                                                                 & \multicolumn{1}{c|}{313.88}     & \multicolumn{1}{c|}{21.56}        & 2469.79      & \multicolumn{1}{c|}{1427.82}    & \multicolumn{1}{c|}{105.12}        & 5765.93      \\ \hline
Ansible                                                          & 2             & 55.44\%                                                                 & \multicolumn{1}{c|}{12411.19}   & \multicolumn{1}{c|}{14.98}        & 62102.16     & \multicolumn{1}{c|}{21847.50}   & \multicolumn{1}{c|}{756.97}        & 83548.89     \\ \hline
Discourse                                                        & 2             & 97.08\%                                                                 & \multicolumn{1}{c|}{26.88}      & \multicolumn{1}{c|}{0.08}         & 540.84       & \multicolumn{1}{c|}{2592.50}    & \multicolumn{1}{c|}{336.03}        & 9848.18      \\ \hline
Hazelcast                                                        & 1             & 15.44\%                                                                 & \multicolumn{1}{c|}{3744.86}    & \multicolumn{1}{c|}{86.65}        & 23836.46     & \multicolumn{1}{c|}{5033.12}    & \multicolumn{1}{c|}{195.82}        & 33223.83     \\ \hline
Kubernetes                                                       & 5             & 45.79\%                                                                 & \multicolumn{1}{c|}{1384.77}    & \multicolumn{1}{c|}{4.40}          & 7165.30      & \multicolumn{1}{c|}{2923.68}    & \multicolumn{1}{c|}{221.73}        & 13023.83     \\ \hline
Node                                                             & 1             & 4.38\%                                                                 & \multicolumn{1}{c|}{933.54}     & \multicolumn{1}{c|}{37.37}        & 8904.92      & \multicolumn{1}{c|}{967.49}     & \multicolumn{1}{c|}{42.47}         & 8924.73      \\ \hline
Odoo                                                             & 2             & 0.69\%                                                                 & \multicolumn{1}{c|}{41737.13}   & \multicolumn{1}{c|}{1413.17}      & 133080.49    & \multicolumn{1}{c|}{41940.02}   & \multicolumn{1}{c|}{1417.77}       & 133685.71    \\ \hline
Rails                                                            & 2             & 31.37\%                                                                 & \multicolumn{1}{c|}{7001.27}    & \multicolumn{1}{c|}{14.85}        & 53244.61     & \multicolumn{1}{c|}{12247.81}   & \multicolumn{1}{c|}{90.90}          & 79046.70     \\ \hline
Salt                                                             & 2             & 26.77\%                                                                 & \multicolumn{1}{c|}{974.20}     & \multicolumn{1}{c|}{74.07}        & 4344.88      & \multicolumn{1}{c|}{2147.26}    & \multicolumn{1}{c|}{270.62}        & 25623.96     \\ \hline
Spark                                                            & 2             & 66.86\%                                                                 & \multicolumn{1}{c|}{1212.80}    & \multicolumn{1}{c|}{23.58}        & 19356.28     & \multicolumn{1}{c|}{5778.25}    & \multicolumn{1}{c|}{258.23}       & 40950.10     \\ \bottomrule
\end{tabular}}
\end{table*}

\textbf{Time to first human response is significantly higher than universal first response time.} We find that humans' first response time in Akka, Ansible, Discourse, Hazelcast, Kubernetes, Salt, and Spark are significantly higher than their universal first response time. For instance, in Discourse, the median time-to-first-human-response in PRs is 336.03 minutes, while its median time-to-universal-first-response is 0.08 minutes. Further analysis on time-to-first-human-response shows that 43\% to 83\% of PRs within the selected projects received their first response from a human within the 8-hour workday.

\subsubsection{What are the statistics of time to human first response in bot-first PRs vs. human-first PRs?}


We formulate this sub-research question to analyze if the human first response time varies in the presence of a bot-first response. In other words, we are interested in investigating if bots correlate with faster first human response in GitHub PRs.

\begin{table*}[]
\centering
\caption{Time-to-first-human-response (TFHR) in human-first PRs vs. bot-first PRs (in minutes). M = mean, Mdn = median, Std = standard deviation. We apply the Mann-Whitney test (ref. Section~\ref{sec:method}) to identify if the difference of TFHR in two types of PRs is statistically significant. Next, we calculate Cliff's d for projects where we observed significant differences and report them in the ``Effect Size'' column. An \NA value in this column means the difference between the two groups is insignificant.}\label{tab:rq12}
\scalebox{1.1}{ 
\begin{tabular}{{l|ccc|ccc|l}}
\toprule
Project &
   &
 \begin{tabular}[c]{@{}c@{}}TFHR in \\  Human-first PRs\end{tabular} &
  \textbf{} &
   &
  \begin{tabular}[c]{@{}c@{}}TFHR in \\ Bot-first PRs\end{tabular} &
   &
  Effect Size \\ \cline{2-7}
\textbf{} &
  \multicolumn{1}{c|}{M} &
  \multicolumn{1}{c|}{Mdn} &
  Std &
  \multicolumn{1}{c|}{M} &
  \multicolumn{1}{c|}{Mdn} &
  Std &
  \textbf{} \\ \hline
Akka       & \multicolumn{1}{c|}{665.59}   & \multicolumn{1}{c|}{20.55}  & 3709.97   & \multicolumn{1}{c|}{2011.60}  & \multicolumn{1}{c|}{347.12}  & 6883.97   & -0.536 (large)\\ \hline
Ansible    & \multicolumn{1}{c|}{22765.17} & \multicolumn{1}{c|}{467.82} & 78940.02  & \multicolumn{1}{c|}{21109.82} & \multicolumn{1}{c|}{1043.92} & 87074.72  & -0.157 (small) \\ \hline
Discourse  & \multicolumn{1}{c|}{901.27}   & \multicolumn{1}{c|}{119.79} & 3045.47   & \multicolumn{1}{c|}{2643.32}  & \multicolumn{1}{c|}{351.31}  & 9976.81   & -0.255 (small) \\ \hline
Hazelcast  & \multicolumn{1}{c|}{4222.61}  & \multicolumn{1}{c|}{148.63} & 25235.57  & \multicolumn{1}{c|}{9471.44}  & \multicolumn{1}{c|}{872.09}  & 60353.27  & -0.260 (small)	 \\ \hline
Kubernetes & \multicolumn{1}{c|}{2311.61}  & \multicolumn{1}{c|}{193.50}  & 7662.89   & \multicolumn{1}{c|}{3587.64}  & \multicolumn{1}{c|}{250.82}  & 17002.81  & -0.050 (negligible)	 \\ \hline
Node       & \multicolumn{1}{c|}{976.31}   & \multicolumn{1}{c|}{42.43}  & 9104.49   & \multicolumn{1}{c|}{775.30}   & \multicolumn{1}{c|}{43.35}   & 2992.13   & \NA	 \\ \hline
Odoo       & \multicolumn{1}{c|}{41959.86} & \multicolumn{1}{c|}{1419.50} & 133445.03 & \multicolumn{1}{c|}{39065.45} & \multicolumn{1}{c|}{1181.40}  & 166105.78 & \NA	 \\ \hline
Rails      & \multicolumn{1}{c|}{10193.79} & \multicolumn{1}{c|}{78.37}  & 64016.28  & \multicolumn{1}{c|}{16741.80} & \multicolumn{1}{c|}{123.62} & 104526.57 & -0.088 (negligible)	 \\ \hline
Salt       & \multicolumn{1}{c|}{1318.59}  & \multicolumn{1}{c|}{234.69} & 5033.05   & \multicolumn{1}{c|}{4414.11}  & \multicolumn{1}{c|}{382.05}  & 48754.77  & -0.161 (small)	 \\ \hline
Spark      & \multicolumn{1}{c|}{2622.37}  & \multicolumn{1}{c|}{38.15}  & 26231.60  & \multicolumn{1}{c|}{7342.80}  & \multicolumn{1}{c|}{503.32} & 46473.89  & -0.576 (large)	 \\ \bottomrule
\end{tabular}
}
\end{table*}


\textbf{The first human response can still be late when a bot generates the first response.} As shown in Table~\ref{tab:rq12}, in 8/10 selected projects, excluding Node and Odoo, the developers in bot-first PRs are taking significantly more time to respond than the developers in human-first PRs. This indicates that the presence of bots may not correlate with a quicker follow-up by humans. Similar as before, the project Odoo, where involvement of bots is low, shows different characteristics. The Node project showed a similar response time in both categories. 



\subsubsection{What are the statistics of the bot-first response time?}


In this sub-research question, we aim to understand the normal bot-first response time in bot-first PRs. 

\begin{table}[t]
\centering
\caption{  Statistics of time-to-bot-first-response across selected projects (in minutes).} \label{tab:rq13}
\small
\begin{tabular}{{lccc}}
\toprule
Project & Median & Mean & Standard Deviation   \\ \hline
Akka             & 22.75           & 44.50         & 254.40                \\ 
Ansible          & 6.50             & 4087.95       & 42339.65      \\ 
Discourse        & 0.08            & 0.61          & 32.61            \\ 
Hazelcast        & 26.94           & 1128.69       & 13583.00       \\ 
Kubernetes       & 0.12            & 379.34        & 6434.30        \\ 
Node             & 0.03            & 0.62          & 9.12               \\ 
Odoo             & 199.35          & 9461.24       & 51213.94            \\ 
Rails            & 0.05            & 16.32         & 824.57                 \\ 
Salt             & 22.10            & 32.11         & 120.28            \\ 
Spark            & 16.12           & 514.00        & 14759.54             \\ \bottomrule
\end{tabular}
\end{table}

\textbf{Bots are present in all selected projects and could respond quickly within a day (median).} The median bot-first response time of half of the projects is lower than 10 minutes, with Node, Rails, Discourse, Kubernetes, and Ansible respectively requiring 0.03, 0.05, 0.08, 0.12, and 6.50 minutes (ref. Table~\ref{tab:rq13}). While the other projects have a median bot-first response time below 27 minutes, Odoo requires a median of 199.35 minutes, which is much slower than other projects. One potential reason is that Odoo only has 0.69\% of its PRs being bot-first PRs.

To understand why bots respond fast in the above-mentioned four projects, we manually checked the bots in their bot-first PRs and found that:
\begin{itemize}
    \item In Discourse, discoursebot \cite{discoursebot} is a CLA bot that asks new PR author to sign the CLA for review right after their PR creation.
    \item In Kubernetes, there five different bots are involved. Among them, k8s-ci-robot \cite{k8s-ci-robot} runs tests, adds labels, and merges code. The involvement of k8s-ci-robot in labeling PRs is noticeable during our inspection, with a quick response. The fejta-bot \cite{fejta-bot} automatically marks issues as stale/rotten and closes them if they do not have any activity for a certain number of days. 
    \item In Rails, rails-bot \cite{rails-bot} is a multitasking bot that assigns reviewers and performs the same operations as fejta-bot. It also adds labels to PRs. 
    \item In Node, nodejs-github-bot \cite{nodejs-github-bot} adds labels and runs tests.
\end{itemize}




\begin{tcolorbox}
\textit{RQ1 Summary: Bots are present in all targeted projects, with a significant percentage (maximum of 97.08\%) of first responses generated by bots.  As such, the universal first response is largely biased by considering responses generated by bots. Despite the quick response of bots, the first human response in bot-first PRs can still be delayed. } 
\end{tcolorbox}

\subsection{\rqtwo } \label{sec:rq2}

\noindent \textbf{Approach:} Zhang et al. showed that, when a pull request has comments, time to first response (comment) plays the most important role in influencing pull request latency, and it explains 58.7\% of the variance in the lifetime of a PR~\cite{zhang2022pull}. In other words, a shorter time-to-universal-first-response is generally associated with a shorter PR lifetime. However, their study does not consider whether a human or a bot made the first comment. We hypothesize that since bots in bot-first PRs tend to respond quickly (as per the results of RQ1), their response times, i.e., the time-to-universal-first-response, may not accurately indicate the remaining processing time for the PR. This is because the developers have yet to become involved in the review task.

To verify our hypothesis and understand the influence of bots on the matter, we constructed a regression model to examine the correlation between time-to-bot-first-response/time-to-first-human-response and PR latency in bot-first PRs. To initiate the process, we utilized the features discussed in Section~\ref{sec:method} and incorporated the feature \textit{botbeforefirst}, indicating whether a bot generated the first response. Following this, we eliminated pull requests, to which humans responded first. Subsequently, we trained a mixed-effects linear regression model based on the \textit{has\_comments=1} approach provided by Zhang et al.~\cite{zhang2022pull}. Finally, we conducted ANOVA Type-II analysis \cite{lmerTest} on the regression outcomes and identified the statistically significant features that account for the variance in the pull request lifetime.

\vspace{0.2cm}

\noindent \textbf{Results:} \textbf{The time-to-first-human-response correlates significantly with PR lifetime, while the time-to-bot-first-response does not.} We found that, in bot-first PRs, the \textit{time-to-bot-first-response} explains \textit{0.33\%} of the variance in the PR lifetime. However, the time of the subsequent first human response, i.e., \textit{time-to-first-human-response} in the same pull request explains \textit{64.70\%} of the variance in the pull request lifetime. 

We conclude that the effect of a bot's first response time in the PR lifetime is negligible, while the subsequent human first response time plays an important role. Future studies on the relationship between the time to first response and the PR lifetime should take into account whether the first response is made by bots, and focus on the human first response instead.

\begin{tcolorbox}
\textit{RQ2 Summary:} The time to first response, when the first response is made by a bot, does not have a significant correlation with the PR lifetime, explaining only 0.33\% of its variance. The time to the subsequent first human response explains 64.70\% of the variance in the PR lifetime.
\end{tcolorbox}

\subsection{\rqthree }

\begin{table*}[]
\centering
\caption{Significance of different features across the studied projects for RQ3. Each value in the significance column represents the number of projects in which statistically significant differences are observed between PRs with longer vs. shorter time-to-first-human-response. (+) means that a long time to response PRs' feature value is often greater than the short time to first response's feature value. (-) represents the opposite. (+,-) shows a mixed relationship, i.e., in some projects, the feature value in the long time-to-response PRs is greater, and in some projects, the opposite pattern is observed. We did not include the category of ``Large'' effect size in this table because no features were found that exhibited such a pattern.} 
\label{sig_features}
\small
\begin{tabular}{{>{\itshape}clccc}}
\toprule
\textup{Dimension} & Features                 & Significance & Small & Medium  \\ \hline
Pull Request       & description\_length               & 5                     & (+) 5          & -                           \\ \cline{2-5}
                   & first\_pr                         & -                     & -              & -                            \\ \cline{2-5}
                   & hash\_tag                         & 2                     & (+) 2          & -                            \\ \cline{2-5}
                   & at\_tag                           & 4                     & (+) 4          & -                            \\ \cline{2-5}
                   & num\_commits\_open                & 4                     & (+) 4          & -                            \\ \cline{2-5}
                   & files\_changed\_open              & 4                     & (+) 4          & -                             \\ \cline{2-5}
                   & src\_churn\_open                  & 5                     & (+) 5          & -                            \\ \cline{2-5}
                   & test\_churn\_open                 & 2                     & (+) 2          & -                             \\ \cline{2-5}
                   & churn\_addition\_open             & 4                     & (+) 4          & -                             \\ \cline{2-5}
                   & churn\_deletion\_open             & 3                     & (+) 3          & -                             \\ \cline{2-5}
                   & commits\_on\_files\_touched\_open & 1                     & (-) 1          & -                            \\ \hline
Project            & team\_size                        & 1                     & (-) 1          & -                             \\ \cline{2-5}
                   & project\_age                      & 5                     & (+,-) 3        & (+) 2                         \\ \cline{2-5}
                   & open\_pr\_num                     & 4                     & (+) 2          & (+) 2                         \\ \cline{2-5}
                   & sloc                             & 3                     & (+) 2          & (+) 1                         \\ \cline{2-5}
                 
                   & perc\_external\_contribs          & 2                     & (-) 2          & -                               \\ \cline{2-5}
                   & integrator\_availability          & -                     & -                           & -          \\  \hline
Developer          & prior\_review\_num                & 1                     & (+) 1          & -                             \\ \cline{2-5}
                   & same\_affiliation                 & 2                    & (+,-) 2              & -                         \\ \cline{2-5}
                   & core\_member                      & 1                     & (-) 1          & -                           \\ \cline{2-5}
                   & prev\_pullreqs                    & 4                     & (-) 4          & -                             \\ \cline{2-5}
                   & social\_strength                  & 4                     & (-) 4          & -                            \\ \cline{2-5}
                   & same\_user                        & 1                     & (-) 1          & -                             \\ \cline{2-5}
                   & followers                         & 1                     & (-) 1          & -                             \\ \bottomrule
\end{tabular}
\end{table*}

To understand the characteristics of PRs with a long time to first response, we have conducted a quantitative and qualitative approach. 

\vspace{0.1cm}
\subsubsection{Quantitative Approach}

We categorize all pull requests into two categories by their time-to-first-response, i.e. long time-to-first-response PRs that take more than one week to receive their first responses, and other PRs. The threshold of one week was determined based on the data distribution and previous works~\cite{maddila2020nudge,kudrjavets2022mining}. Next, we perform statistical tests comparing the feature values of these two groups of PRs.

Table~\ref{sig_features} summarizes the significance of different features across the studied projects. We consider a feature significant if its difference between PRs with long time to first response and PRs with short time to response is both statistically significant and practically significant (i.e., the effect size is small, medium, or large) in at least one project. A given cell in the table counts the number of projects in which a given significance is observed.



\vspace{0.1cm}


\noindent \textbf{Results:} \textbf{In approximately half of the projects, nine features demonstrate consistent significance.} As shown in Table~\ref{sig_features}, these nine features are \textit{description\_length, num\_commits\_open, files\_changed\_open, src\_churn\_open, churn\_addition\_open, at\_tag, open\_pr\_num, prev\_pullreqs, and social\_strength}. Five out of the nine significant features are proposed to measure the complexity of a PR at the opening time from different aspects. \textbf{Specifically, PRs that experience a prolonged first human response time often have a longer description and more complex code change at the PR opening time.} 

We also observe that in four projects, PRs take a longer time to receive the first human response as they experience a higher workload (increase in open pull requests~\cite{khatoonabadi2021wasted} - high \textit{open\_pr\_num}). Two features, i.e., \textit{project\_age} and \textit{same\_affiliation}, have both positive and negative correlations with time-to-first-human-response, depending on the specific projects.





\textbf{The contributors of PRs with long time-to-first-human-response tend to have less experience and be less communicative.} As shown in Table~\ref{sig_features}, PRs with authors having less experience (\textit{prev\_pullreqs}) tend to have a longer time to receive first human response. Furthermore, lack of interaction with the project members correlates with a longer first human response time (\textit{social\_strength}) in 4 projects. 

\vspace{0.2cm}
\subsubsection{Qualitative Approach}
We  manually investigated 100 PRs (randomly sampled 10 from each project) that waited for their first response for more than one week, in order to find the potential reasons behind the delay. To mitigate subjective bias, two annotators independently performed open coding on the first human response, read the whole PR containing the first response, labeled it, then discussed until they agreed on the final labels. 

\noindent \textbf{Results:} \textbf{We found that none of the reviewers explicitly explained why they were late for the first response.} In 23\% of the examined PRs, there are some later comments in the PRs indicating potential reasons. The results are similar to the findings from the long time-to-first-response samples manually checked in the preliminary study (see Section~\ref{sec:preliresults}). Specifically, 15\% of the PRs are incomplete when submitted. Therefore, the delay may be due to waiting for the PR completion. 7\% of the PRs are rejected or suspended for reasons including targeting an older version of the software, issues being fixed in another PR, and the PR being a proposed draft not intended to be merged. In one PR, the reviewer mentioned that \textit{``we are currently in a code freeze period for new features until 2.4 ships, which will be around next week or so. Once that is done, we can pick this up.''}, thus code freeze may be the reason for the delay of the first response.



\vspace{0.1cm}
\begin{tcolorbox}
\textit{RQ3 Summary:} Complex PRs with lengthy descriptions and inexperienced contributors with less communicative attitudes correlate with a longer time to receive the first human response. Nine features from three dimensions, i.e., PR, project, and developer, demonstrate consistently significant differences between long and short time-to-first-human-response in around half of the projects.
\end{tcolorbox}
\vspace{0.1cm}

\subsection{\rqfour }  \label{sec:rq4}

\noindent \textbf{Approach:} To understand the relationship between future contributions of a pull request author and the time to a first human response on the PRs of that author, especially for newcomers, we split our collected pull requests into two groups based on the universal human response time, i.e., PRs who received their first human response within one day and PRs who received their first human response after more than one day. Next, we further split PRs based on whether or not a newcomer created the PR, i.e., the PR is the first one created by the author. We report the ratio of developers who continue to contribute in the future based on the two categories.

\vspace{0.1cm}

\noindent \textbf{Results:} \textbf{For all ten considered projects, newcomers who got their first response from humans earlier had a higher probability of contributing again to a given project.} As shown in Table~\ref{tab:rq41}, though newcomers in different projects have a different likelihood to continue contributing, the ratio of continuing newcomers who received their first response within one day is 2.63\%-37.44\% higher than those who received their first response late, with a median of 15.14\%. 

\begin{table}[t]
\centering
\caption{Ratios of newcomers continuing to contribute depending on quicker or slower first response time on their first contribution.} \label{tab:rq41}
\scriptsize
\begin{tabular}{{lllll}}
\toprule
\multirow{2}{*}{Project}&\multirow{2}{*}{\#PRs} &\multicolumn{2}{c}{\% PR Author Continuing}&\multirow{2}{*}{Delta (\%)}\\\cline{3-4}
& & \vtop{\hbox{\strut PRs with}\hbox{\strut $<=1$ day TFHR}} & \vtop{\hbox{\strut PRs with}\hbox{\strut $>1$ day TFHR}}& \\ \hline
Akka         & 687              & 44.77\%           & 38.35\%  & +16.74\%                               \\ \hline
Ansible      & 3,367 & 45.85\% & 39.82\%     &    +15.14\%                 \\\hline
Discourse    & 824 & 40.72\% & 32.49\%   &      +25.33\%                           \\ \hline
Hazelcast    & 186 & 51.54\% & 37.50\%   &      +37.44\%                          \\ \hline
Kubernetes   & 1,319 & 53.29\% & 40.07\%   &    +32.99\%                       \\ \hline
Node         & 1,455 & 38.03\% & 35.71\%   &        +6.50\%                 \\ \hline
Odoo         & 794 & 62.06\% & 60.47\%   &            +2.63\%             \\ \hline
Rails        & 3,146 & 42.25\% & 39.02\%        &    +8.28\%                     \\ \hline
Salt         & 1,378 & 52.69\% & 48.06\%     &    +15.14\%                      \\ \hline
Spark        & 2,040 & 49.19\% & 43.29\%    &               +9.63\%        \\ \bottomrule
\end{tabular}

\end{table}

To investigate if the above observation is also true for existing contributors, we perform similar analysis and focus on pull requests created by non-first time PR contributors. Table~\ref{tab:rq42} shows the results. Out of the ten considered projects, nine of them show a similar pattern, i.e., PR authors who received their first human response within one day were more likely to continue their contributions, except for the \textit{Kubernetes} project. The improvement ratio ranges from 0.06\% to 9.48\%, with a median of 2.25\%. 

\begin{table}[t]
\centering
\caption{Ratios of existing contributors continuing to contribute depending on quicker or slower first response time on their first contribution.} \label{tab:rq42}
\scriptsize
\begin{tabular}{{lllll}}
\toprule
\multirow{2}{*}{Project}&\multirow{2}{*}{\#PRs} &\multicolumn{2}{c}{\% PR Author Continuing}&\multirow{2}{*}{Delta (\%)}\\\cline{3-4}
& & \vtop{\hbox{\strut PRs with}\hbox{\strut $<=1$ day TFHR}} & \vtop{\hbox{\strut PRs with}\hbox{\strut $>1$ day TFHR}}& \\ \hline
Akka         & 5,891 & 95.26\% & 93.44\%    &      +1.95\%                       \\ \hline
Ansible      & 12,020 & 89.23\% & 85.75\%   &           +4.06\%                    \\ \hline
Discourse    & 3,495 & 86.85\% & 79.33\%     &      +9.48\%                       \\ \hline
Hazelcast    & 4,127 & 98.72\% & 98.21\%    &        +0.52\%                     \\ \hline
Kubernetes   & 17,344 & 95.69\% & 95.92\%   &         -0.24\%                       \\ \hline
Node         & 6,963 & 93.20\% & 91.45\%    &           +1.91\%                      \\ \hline
Odoo         & 8,982 & 94.64\% & 92.26\%   &          +2.56\%                  \\ \hline
Rails        & 9,561 & 85.75\% & 83.51\%   &          +2.68\%                      \\ \hline
Salt         & 8,749 & 91.01\% & 88.32\%    &          +3.04\%                      \\ \hline
Spark        & 18,766 & 94.88\% & 94.82\%   &           +0.06\%                   \\ \bottomrule
\end{tabular}

\end{table}

\vspace{0.1cm}
\begin{tcolorbox}
\textit{RQ4 Summary: In all the considered ten projects, the ratio of newcomers continuing their PR contribution is 2.63\%-37.44\% higher when the newcomers received their first human response within one day, compared to more than one day.} 
\end{tcolorbox}
\vspace{0.2cm}


%% file: sections/discuss.tex


\subsection{Implications for research and practice}
We summarize the implications for different stakeholders below by combining the results from our preliminary study and four RQs.

\vspace{0.1cm}
\noindent \textbf{(1) The presence of bots in GitHub pull requests would impact the universal first response time in PR.} Therefore, researchers need to distinguish between bot response and human response and should not consider only the universal first response. Recent studies have shown that bots or development supporting tools are commonly adopted in modern OSS projects~\cite{wessel2020effects,wang2022specialized}. We observe the same in the ten selected projects. Specifically, we found that bots generate 0.69\% to 97.08\% of the first response in our target PRs. Those bots usually respond within 10 minutes after the PR creation, which is much faster than humans. Without separating the bot-first PRs from human-first PRs, researchers would wrongly interpret the statistics, such as the median and mean time to first response of a PR (see RQ1). 

\vspace{0.1cm}
\noindent \textbf{(2) The first human response time is highly correlated with PR lifetime. On the other hand, the time to bot generated universal first response does not have a significant correlation with the PR lifetime.} Therefore, researchers and developers should consider first human response as an essential estimation factor for PR lifetime. We build regression models to investigate the relationship between bot first response time, first human response time, and PR lifetime in bot-first PRs by considering important features reported in a recent study on PR latency~\cite{zhang2022pull2}. Our results show that bot first response time explains 0.33\% of the variance in the PR lifetime, while first human response time explains 64.7\% of it. 

\vspace{0.1cm}
\noindent \textbf{(3) Bot adoption does not necessarily correlate with faster human first response in PRs. Moreover, the ratio of newcomers continuing their PR contribution is higher (as high as 37.44\%) when they received their first human response earlier.} Therefore, developers should still try to respond to the PR author as soon as possible and use tools to facilitate the PR management process and avoid delay in the first stage. In RQ4, we find that in all our analyzed projects, newcomers would have a (2.63\%-37.44\%) higher ratio to contribute another PR if they received their first human response within one day, than those who received their first response in more than one day. Such a pattern is minor for existing PR contributors. Surprisingly, we observe that human-first response times in bot-first PRs are statistically significantly slower than those in human-first PRs in most selected projects (ref. Section~\ref{sec:rq1}). This indicates that using a bot might not generate pressure for project maintainers to follow up rapidly. Future research could be conducted to understand bots' role in pull request latency analysis.


%% file: sections/threats.tex
This work is based on a dataset extended from the PR latency benchmark provided by Zhang et al.~\cite{zhang2022pull2}. This means, this paper inherits all the threats introduced by their data collection process. As we extend the benchmark dataset by collecting new pull request information from GitHub, the data collection process might introduce some noise, though we use the official GitHub v4 API. Furthermore, this paper also suffers from the following internal and external threats. 

\vspace{0.1cm}
\noindent \textbf{Threats to internal validity:} We identified bots in GitHub PRs leveraging a tool named BoDeGHa, which is the most cited bot detection approach until now. However, we observed many false positives while checking its results, e.g., core developers are identified as bots. We then heuristically examined the predicted results and refined the bot list. However, we might miss some real bots. To mitigate this risk, we examined 1,000 randomly sampled PRs from our dataset and manually evaluated if their information aligns with that shown on the corresponding GitHub pages. We only observed 0.2\% inconsistency (ref. Section~\ref{sec:method}). In RQ4, we only check the ratio of developers who continue to create PRs, as a proxy for the future activities of PR authors. In practice, there could be other types of contributions. Also, the extended 1-year PRs might not capture the future activities of some PR authors. 

We define response as PR comment or code review comment in this work. However, there are other events in pull requests. Those events can be considered as another type of response. Therefore, our findings might not be generalized if events are included as responses. The last internal threat came from the PR filtering process. Since this work focuses on the time to first response, especially the first human response, we remove all PRs that do not have any response (comment) for the preliminary study, and we exclude all PRs without human response for RQ1-4. Thus, our findings may not be generalized to those PRs. We did not consider the presence of mixed accounts in the study, however, we did not encounter any indication of their presence when we manually examined 1000 PRs (see ref. Section~\ref{sec:method}).


\vspace{0.1cm}
\noindent \textbf{Threats to external validity:} We mainly focused on ten popular projects with the richest historical PR data. Although the selected projects are in different domains, our findings may not generalize to all open-source projects on GitHub. Future studies should expand the analysis to explore more diverse projects. Like Zhang et al.~\cite{zhang2022pull2}, we adopt the Mixed-Effect Linear Regression model to investigate the relationship between time to first response and PR lifetime (ref. Section~\ref{sec:rq2}). However, different regression models may return different results.